%% file: article8641-11.tex
\documentclass[a4paper]{spie}  
\usepackage[]{graphicx}
\usepackage{amssymb,amsmath,psfrag,url,enumerate,caption}
\usepackage[german,english]{babel}
\include{MatheDef}

\title{Numerical analysis of nanostructures \\for enhanced light extraction from OLEDs} 

\author{Lin Zschiedrich\supit{a}, Horst J.~Greiner\supit{b}, Sven Burger\supit{a,c}, and Frank Schmidt\supit{a,c} 
\skiplinehalf
\supit{a}JCMwave GmbH, Bolivarallee 22, 14050 Berlin, Germany\\
\supit{b}Philips Research Aachen, Wei{\ss}hausstra{\ss}e 2, 52066 Aachen, Germany  \\
\supit{c}Zuse Institute Berlin, Takustra{\ss}e 7, 14195 Berlin, Germany  
}

\authorinfo{Further author information: (Send correspondence to Lin Zschiedrich, E-mail: lin.zschiedrich@jcmwave.com\,.)}

\begin{document} 
  \maketitle 

\noindent
This paper will be published in Proc.~SPIE Vol. {\bf 8641}
(2013) 86410B, ({\it Light-Emitting Diodes: Materials, Devices, 
and Applications for Solid State Lighting XVII}, DOI: 10.1117/12.2001132), 
and is made available 
as an electronic preprint with permission of SPIE. 
One print or electronic copy may be made for personal use only. 
Systematic or multiple reproduction, distribution to multiple 
locations via electronic or other means, duplication of any 
material in this paper for a fee or for commercial purposes, 
or modification of the content of the paper are prohibited.
Please see original paper for images at higher resolution. 

\begin{abstract}
Nanostructures, like periodic arrays of scatters or low-index gratings, are used to improve the light outcoupling from organic light-emitting diodes (OLED). In order to optimize geometrical and material properties of such structures, simulations of the outcoupling process are very helpful. The finite element method is best suited for an accurate discretization of the geometry and the singular-like field profile within the structured layer and the emitting layer. However, a finite  element simulation of the overall OLED stack is often beyond available computer resources. The main focus of this paper is the simulation of a {\em single} dipole source embedded into a twofold infinitely periodic OLED structure. To overcome the numerical burden we apply the Floquet transform, so that the computational domain reduces to the unit cell. The relevant outcoupling data are then gained by inverse Flouqet transforming. This step requires a careful numerical treatment as reported in this paper.  
\end{abstract}



\section{Introduction}
\label{sec:intro}  
\begin{figure}
 \psfrag{x}{$x$}
 \psfrag{y}{$y$}
 \psfrag{z}{$z$}
 \psfrag{R}{$R$}
 \psfrag{Scatterers}{scatterers}
 \psfrag{Emitters}{emitters}
 \psfrag{Substrate}{substrate}
 \psfrag{Organic}{organic}
 \psfrag{Cathode}{cathode}
 \psfrag{Anode}{anode}
 \psfrag{Substrate}{substrate}
 \psfrag{Superstrate}{superstrate}
  \begin{center}
      \includegraphics[width=10cm]{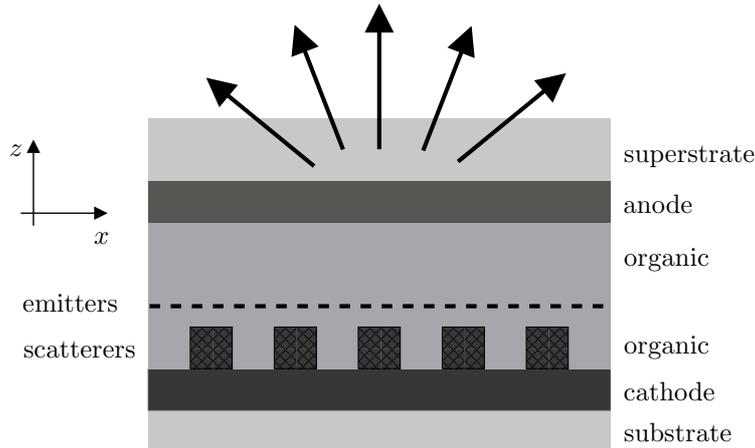}
  \end{center}
  \caption{\label{Fig:OledStack} Simplified OLED structure. Light is generated in a slim emitter layer and radiated into the upper half-space. Tiny scatterers are used to increase the light extraction efficiency. 
 }
\end{figure}

Figure~\ref{Fig:OledStack} shows a simplified OLED structure.
It essentially consists of a layered medium stack. Light is generated in a slim emitting layer within the organic semiconductor material. Often, the metallic cathode also serves as an optical back reflector. The goal is to extract as much of the generated light as possible into the superstrate. Full extraction is inhibited due to the presence of lossy materials and by the trapping of light into waveguide modes by total reflection. The waveguide modes travel in horizontal direction and are therefore lost for emission. Scattering particles are commonly used to disrupt the propagation of the waveguide modes; the light, gradually scattered by the particles, can then leave the OLED device. Unfortunately, periodically arranged scatterers are in general not able to guarantee full light extraction even not for transparent materials: Light is still trapped in Bloch modes (except for frequencies within the {\em band gap} of the twofold photonic crystals). A proper design of the periodic arrangement is therefore of major importance for an efficient OLED.                

The simulation of light extraction properties is a numerical challenge for the following reasons:
\begin{enumerate}
\item A wavelength scan over the entire visible spectrum is needed.
\item Many light emitters have to be simulated at different positions within the structured OLED. 
\item Metals gives rise to the presence of plasmons with singular field profiles near metallic edges or corners. \label{matmod}
\item Realistic material data are only given experimentally. A numerical dispersion model can be costly to implement. 
\item The computational domain must be sufficiently large to suppress numerical truncation errors.  
\end{enumerate}
In this paper we focus on the finite element method (FEM) in the frequency domain. This method allows for an efficient and accurate discretization of the geometry as well as an automatic mesh adaption for an accurate resolution of the highly nonuniform field profiles.  Since we apply FEM in frequency domain each wavelength is treated separately. However, multiple dipole source positions can be computed efficiently in one sweep. This is because the finite element method is chiefly limited by the direct sparse matrix solver, which can be re-used for different source terms. 

In this paper we discuss a method which allows to simulate an isolated source embedded in a twofold periodic arrangement without the need to use a large computational domain. By means of the Floquet transform, see Kuchment in Gao et al.~\cite[pp. 207]{Bao:87}, the original problem posed on the entire periodic space is mapped to a bundle of Bloch-periodic problems posed on the unit cell of the periodic structure. These Bloch-periodic problems can be solved with a tremendous reduction of memory requirements. The price we have to pay is the inverse Floquet transform which is an integration over the Brillouin zone of the reciprocal lattice space. Using adaptive integration techniques together with a straightforward parallelization we show that this can be done with reasonable numerical costs. The idea to numerically employ the inverse Floquet transform goes back to Wilcox, Botten, McPhedran et al~\cite{Wilcox:05a}. There, the motivation was the computation of defect modes in photonic crystals. It has been shown that the inverse Floquet transform is still numerically feasible even close to the band-edge of the photonic crystal. 

The paper is organized as follows: In Section~\ref{Sec:MaxEmit} we settle the basic concepts for modelling light extraction from a light emitting diode. Then we explain how the finite element method accurately deals with singular sources such as point dipoles (Section~\ref{Sec:FEMDip}). Section~\ref{Sec:PerGeo} introduces the Floquet-transform techniques. The final section covers numerical concepts and examples.

\section{Maxwell's equations: light emission model}
\label{Sec:MaxEmit}
We consider Maxwell's equations in the frequency domain. That is, we assume a time-harmonic dependency of the electromagnetic field, i.e.,
\begin{eqnarray*}
\VField{E}(\pvec{r}, t) & = & \Re \left( \hat{\VField{E}}(\pvec{r}, \omega) e^{-i\omega
  t }\right), \\
\VField{H}(\pvec{r}, t) & = & \Re \left( \hat{\VField{H}}(\pvec{r}, \omega) e^{-i\omega
  t }\right),
\end{eqnarray*} 
and accordingly for the source current $\VField{J}.$ The constitutive relations are given in the form
\begin{subequations}
\label{ConstitutiveEqn}
\begin{eqnarray}
\label{ConstitutiveEqn1}
\hat{\VField{D}}(\pvec{r}, \omega) & = & \varepsilon(\pvec{r}, \omega) \,
\hat{\VField{E}}(\pvec{r}, \omega), \\ 
\label{ConstitutiveEqn2}
\hat{\VField{B}}(\pvec{r}, \omega) & = & \mu(\pvec{r}, \omega) \,
\hat{\VField{H}}(\pvec{r}, \omega), \\ 
\label{ConstitutiveEqn3}
\hat{\VField{J}}(\pvec{r}, \omega) & = & \sigma(\pvec{r}, \omega) \,
\hat{\VField{E}}(\pvec{r}, \omega) + \hat{\VField{J}}_{i}(\pvec{r}, \omega), 
\end{eqnarray}
\end{subequations}
with the material tensors permittivity, $\varepsilon$, permeablity, $\mu$, and conductivity, $\sigma$. 
$\hat{\VField{J}}_{i}(\pvec{r}, \omega)$ is the impressed current density. For simplicity we will henceforth drop the hats, will use $\VField{J}$ for the the impressed source, and we introduce the complex permittivity tensor $\varepsilon = \varepsilon+i\sigma/\omega.$ Then, Maxwell's equations can be cast into a second order form for the electric field, 
\begin{eqnarray}
\label{ETHMax}
\curl \mu^{-1} \curl \VField{E}(\pvec{r})-\omega^2 \varepsilon \VField{E}(\pvec{r}) & = & i \omega
\VField{J}(\pvec{r}).
\end{eqnarray}
In an OLED simulation, the major quantity of interest is the extraction efficiency. This is defined as the quotient of the power $P_{\mathrm{rad}},$ radiated into the superstrate, and the total emitted power $P_\mathrm{tot}$ of the source, 
\begin{align}
\label{Eqn:TotEmit}
\eta_{\mathrm{rad}} = \frac{P_{\mathrm{rad}}}{P_\mathrm{tot}}.
\end{align}

\begin{figure}
  \begin{center}
      \includegraphics[width=8cm]{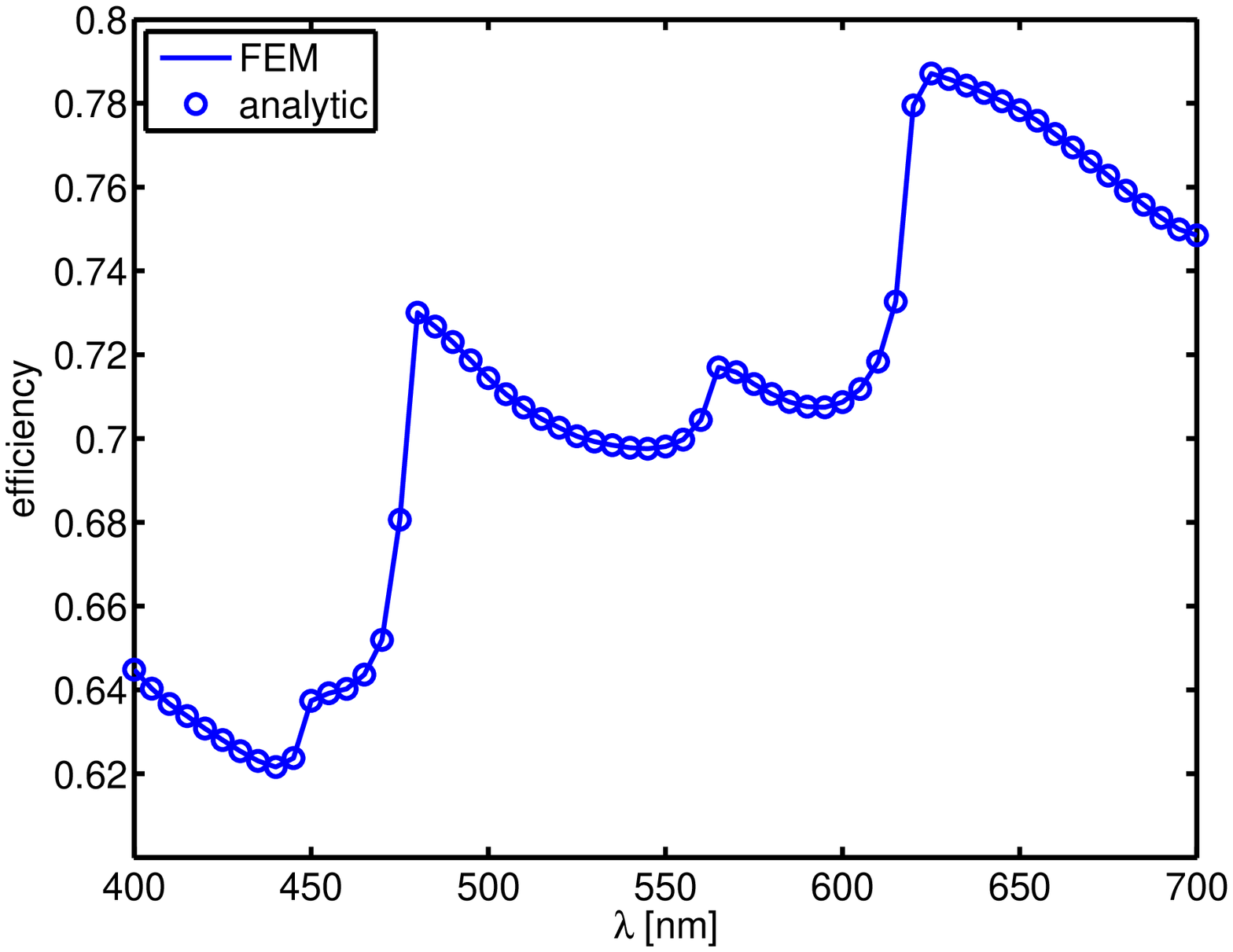}
      \includegraphics[width=8cm]{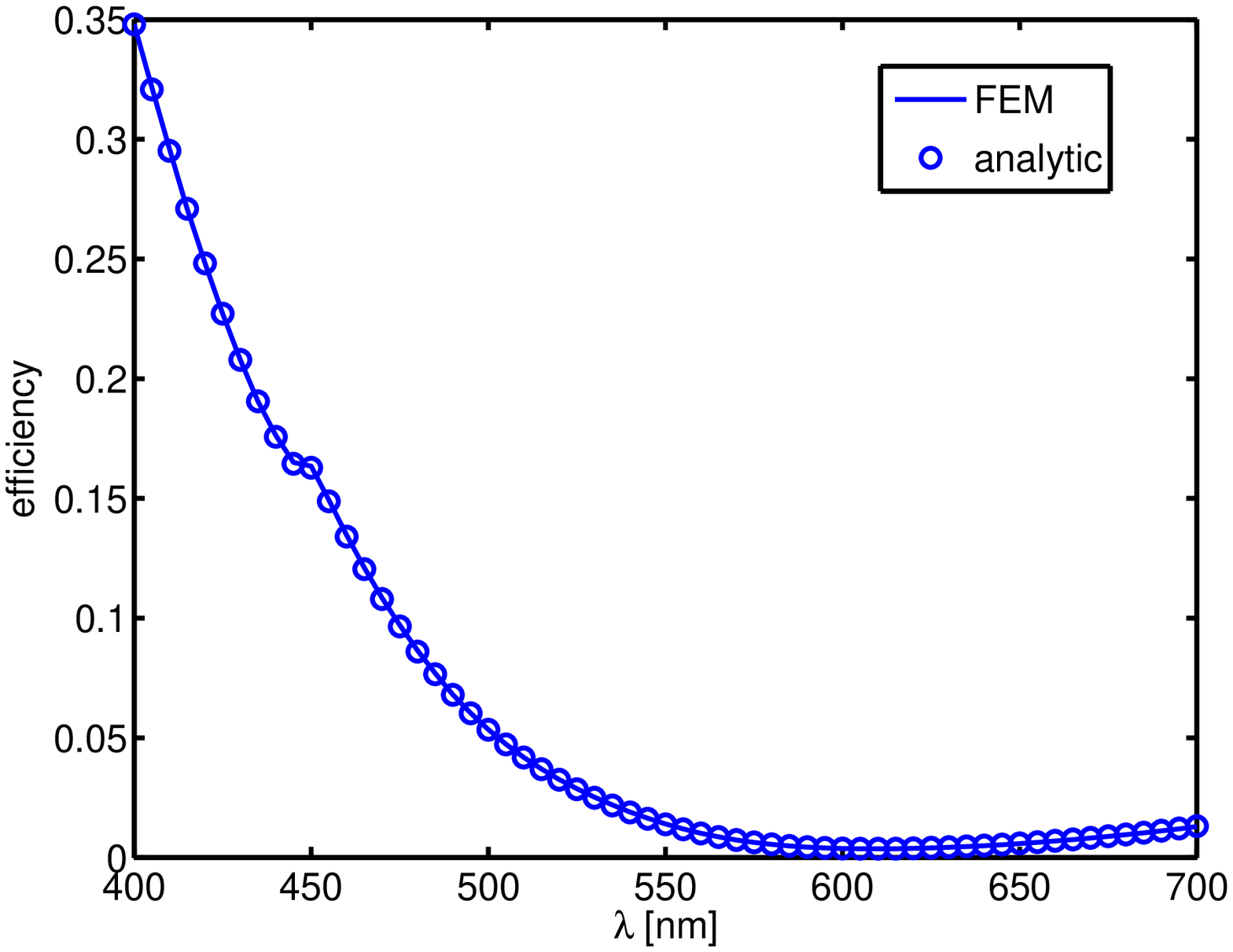}
  \end{center}
  \caption{\label{Fig:PlanarOledEfficiency} Computed efficiencies for a planar OLED stack. Left: horizontally oriented dipole. Right: vertically oriented dipole. The lower substrate consists of an infinite silver cathode (Drude model) followed by layers ($n$: refractive index $d$:thickness) $n_1=1.75, d_1=80\mathrm{nm}$; $n_2=1.75, d_2=100\mathrm{nm}$; $n_3=1.8, d_3=200\mathrm{nm}$; $n_4=1.9, d_3=600\mathrm{nm}$, and an glass superstrate with $n=1.5.$ The dipoles are placed between the first and second layer. This non-trivial example demonstrates the fidelity of the FEM approach. 
 }
\end{figure}

More precisely, the radiated power is the integrated power flux through an infinitely far upper hemisphere, that is,
\begin{align*}
\lim_{R\rightarrow \infty} \int_{S_{R, +}} \frac{1}{2} \Re \left( \overline{\VField{E}} \times {\VField{H}} \right) \cdot \pvec{n}\, \dd{S} = \lim_{R\rightarrow \infty} \int_{S_{R, +}} \frac{1}{2} \sqrt{\frac{\varepsilon}{\mu}} \left|\VField{E} \right |^2 \, \dd{S},
\end{align*}
where $S_{R, +}$ is the upper hemisphere with radius $R$ and $n$ is normal vector. The last equality holds true, because the far field satisfies the Silver-M\"uller radiation condition, see Monk~\cite[p. 226]{Monk:03a}. 

To compute the total emitted power $P_\mathrm{tot}$ we regard a domain $\Omega$ containing the source. Then, the emitted power is the energy lost in this domain plus the net power flux across the boundary $\mathrm{\Omega},$
\begin{align*}
P_\mathrm{tot} = \int_{\partial \omega} \frac{1}{2}  \Re \left( \overline{\VField{E}} \times {\VField{H}} \right)\;\dd{S}+\int_\Omega \frac{1}{2} \overline {\VField{E}} \cdot \sigma \VField{E} \, \dd{V}.
\end{align*}
This expression can be simplified to a linear functional for the electric field. To see this, we multiply Maxwell's equations~\eqref{ETHMax} with $\overline{\VField{E}}$ and integrate over $\Omega$:
\begin{eqnarray*}
\int_{\Omega} \overline {\VField{E}}  \cdot \curl \mu^{-1} \curl \VField{E}-\omega^2 \overline {\VField{E}}  \cdot \varepsilon \VField{E}\,\dd{V} = & 
i \omega \int_{\Omega} \overline {\VField{E}}  \cdot \VField{J}\,\dd{V}.
\end{eqnarray*}
Partially integrating yields
\begin{eqnarray*}
\int_{\Omega}\curl \overline   {\VField{E}}\cdot \mu^{-1} \curl \VField{E}-\omega^2 \overline   {\VField{E}} \cdot \varepsilon \VField{E}\,\dd{V}-\phantom{xx} \\
\int_{\partial \Omega}\underbrace{ \left ( \overline {\VField{E}} \times \mu^{-1} \curl \VField{E}\right) }_{=i\omega  \overline {\VField{E}}\times \VField{H}}  \cdot \pvec{n}\;\dd{S}= &i \omega \int_{\Omega} \overline {\VField{E}}  \cdot \VField{J}\,\dd{V}.
\end{eqnarray*}
Recalling that $\Im \left( \varepsilon \right ) = \sigma/\omega$ and taking the imaginary part on both sides gives
\begin{align*}
P_\mathrm{tot} = -\frac{1}{2}  \int_{\Omega}  \Re \left ( \overline   {\VField{E}}\cdot  \VField{J} \right) \,\dd{V}.
\end{align*}
\section{Modelling dipole sources with finite elements}
\label{Sec:FEMDip}
For a dipole source at position $\pvec{r}'$ the impressed electric current is modeled as a delta distribution, $\VField{J}(\pvec{r})=\pvec{p}\delta(\pvec{r}-\pvec{r}')$ with given dipole moment $\pvec{p}.$ 
The regularity of $\VField{E}$ is poor and a direct finite element discretization of $\VField{E}$
suffers from a slow convergence. To cure this we use the subtraction approach, see Awada et al.~\cite{Awada:97a}, Wolters~\cite{Wolters:03a} and Zschiedrich~\cite{Zschiedrich:07a}. 

The idea behind the subtraction approach is to determine an analytically available singular field $\VField{E}_{s}$ which already comprises the singular part of $\VField{E}$ at the dipole position. A natural candidate is the homogeneous Green's function $\VField{E}_{s}$, that is
\begin{align*}
\curl \mu^{-1}_d \curl \VField{E}_{s}(\pvec{r})-\omega^2 \varepsilon_d \VField{E}_{s}(\pvec{r}) = i\omega \pvec{p}\delta(\pvec{r}-\pvec{r}'),
\end{align*}
with a constant material background  $\epsilon_d = \epsilon(\pvec{r}')$ and $\mu_d=\mu(\pvec{r}')$ as given at the dipole position. 
Now, we split the field $\VField{E}$ into the singular field $\VField{E}_{s}$
and a correction field $\VField{E}_{c}$ that is $\VField{E} = \VField{E}_{s}+\VField{E}_{c}.$  Inserting into Maxwell's equations~\eqref{ETHMax} yields 
\begin{align*}
 \curl \mu^{-1} \curl \left(\VField{E}_{s}+\VField{E}_{c} \right)(\pvec{r})-\omega^2 \varepsilon \left(\VField{E}_{s}+\VField{E}_{c} \right)(\pvec{r})  = & { }   \\
\curl \mu^{-1} \curl \VField{E}_{c} (\pvec{r})-\omega^2 \varepsilon \VField{E}_{c} (\pvec{r})+ 
\curl \mu^{-1} \curl \VField{E}_{s} (\pvec{r})-\omega^2 \varepsilon \VField{E}_{s} (\pvec{r}) = & {}  \\
\curl \mu^{-1} \curl \VField{E}_{c} (\pvec{r})-\omega^2 \varepsilon \VField{E}_{c} (\pvec{r})+ 
\curl (\mu^{-1}-\mu^{-1}_d) \curl \VField{E}_{s} (\pvec{r})-\omega^2 (\varepsilon-\varepsilon_d)\VField{E}_{s} (\pvec{r})  + \phantom{xxxx}  \\
\underbrace{\curl \mu^{-1}_d \curl \VField{E}_{s} (\pvec{r})-\omega^2 \varepsilon_d +\VField{E}_{s} (\pvec{r})}_{=i \omega \pvec{p}\delta(\pvec{r}-\pvec{r}')} = & i \omega \pvec{p}\delta(\pvec{r}-\pvec{r}')\,. \phantom{xx}
\end{align*}
Hence, as desired, the singular source terms on both sides cancel out.  
When rearranging the terms we end up with Maxwell's equations for the correction field $\VField{E}_{c}$ only,  
\begin{align*}
{} & \curl \mu^{-1} \curl \VField{E}_{c} (\pvec{r})-\omega^2 \varepsilon \VField{E}_{c} (\pvec{r}) = \\
{} & \phantom{xxxx} - \curl (\mu^{-1}-\mu^{-1}_d) \curl \VField{E}_{s}+\omega^2 (\epsilon-
\epsilon_d)\VField{E}_{s}.
\end{align*}
The analytically given right hand side is equal to zero in a vicinity of the dipole position. Hence, this equation for the correction field $\VField{E}_{c}$ is well suited for an accurate discretization with finite elements. (Even a jump in the permeability is allowed, as in the variational form the most left $\curl-$ operator on the right hand side can be applied on the test function by partial integration.) 

\subsubsection*{Remark}
For simplicity we have assumed so far that the material background is homogeneous in a vicinity of the dipole. The case of a dipole placed on or near a material interface can be treated in the same way: As the singular field $\VField{E}_{s}$, one the chooses the solution of the dipole source for a two layer material, which is available quasi-analytically, see Paulus~\cite{Paulus:00a}.

It remains to evaluate the radiation efficiency $\eta_{\mathrm{rad}}$ in an accurate manner for the dipole case. The total radiated power can be computed as usual from the far field data. But the expression~\eqref{Eqn:TotEmit} for the total emitted power calls for a delicate mathematical justification as it involves the integral of the delta distribution with a singular field. Fortunately, one can show that this expression is properly defined when the dipole is placed in a lossless background. Then, the total emitted power is given by
\begin{align}
\label{Eqn:DipMom}
P_\mathrm{tot} = -\frac{1}{2} \Re \left ( \overline {\VField{E}}(\pvec{r}') \cdot  \pvec{p} \right ) \cdot 
\end{align}
Again, we evaluate the expression by using the splitting $\VField{E} = \VField{E}_{s}+\VField{E}_{c}.$ $\Re \left ( \overline {\VField{E}_s}(\pvec{r}') \cdot  \pvec{p} \right )$ is analytically available, whereas $\VField{E}_{c}$ is smooth and can be evaluated within the finite element framework with high precision. 

To validate the FEM method we compare the results for a planar OLED stack with the quasi-analytic solution as obtained by Fourier expansion techniques, c.f., Paulus et al~\cite{Paulus:00a}. This is a non-trivial test for the FEM approach, since only the homogeneous dipole solution was used as the singular field. Figure~\ref{Fig:PlanarOledEfficiency} shows a great agreement of the 3D numerical solution (FEM) with the exact (analytic) result. 

\section{Periodic geometries}
\label{Sec:PerGeo}

\begin{figure}
  \begin{center}
\begin{minipage}{0.4\textwidth}
      \includegraphics[width=7.0cm]{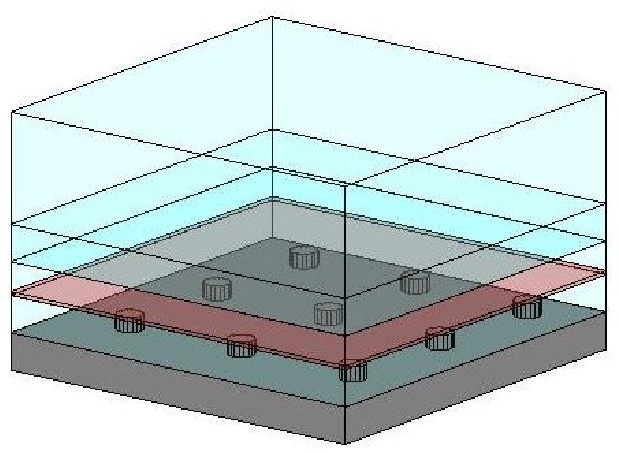}
\end{minipage}
\hspace{1cm}
\begin{minipage}{0.4\textwidth}
\begin{tabular}[t]{rr}
thickness $[\mathrm{nm}]$ & refractive index \\
\hline
$\inf$ & $1.5$ \\
$100$ & $1.8$ \\
$100$ & $1.75$ \\
$80$ & $1.75$ \\
$100$ & Ag (Drude) \\
$\inf$ & $1.5$ \\
\end{tabular}
\end{minipage}
  \end{center}
  \caption{\label{Fig:Geo3D} An OLED stack with an periodically structured cathode. Only finite layers are plotted. The material stack is given in the table. The dipoles are placed between the second and third layer (red plane)}  
\end{figure}

We now address the case of a twofold periodically structured device, that is 
\begin{align*}
\varepsilon(\pvec{r}+\pvec{a}_{1/2}) = &\varepsilon(\pvec{r}), \\
\mu(\pvec{r}+\pvec{a}_{1/2}) = &\mu(\pvec{r}),
\end{align*}
with grid vectors $\pvec{a}_{1},$ $\pvec{a}_{2}$ in the $xy-$plane. 

In the following, let $\pvec{l} = [l_1;l_2] \in \nnum^2$ denote an integer vector and we use the notation $\pvec{a} = [\pvec{a}_1,\,\pvec{a}_2]$ (in matlab style).  We call a source field {\em Bloch-periodic}, when it satisfies 
\begin{align*}
\VField{J}(\pvec{r}+\pvec{a}  \cdot \pvec{l}) = e^{i\pvec{k}_\mathrm{B}^\mathrm{T} \pvec{a} \cdot \pvec{l}} \VField{J}(\pvec{r})
\end{align*}
with a Bloch-phase vector $\pvec{k}_\mathrm{B} \in \rnum^2.$ The corresponding electric field is Bloch-periodic as well. This allows to restrict the computation onto an unit cell by imposing Bloch-periodic boundary conditions. 

However, the main focus of this paper is the simulation of a {\em single} dipole source embedded into a periodic arrangement. Since this single dipole source is not Bloch-periodic, the entire space is in principle needed as computational domain. However, in the sequel we will explain that the computational domain can be reduced to the unit cell by using the Floquet transform. For any sufficiently decaying source term we perform the {\em Floquet} transform on $\VField{J}:$      
\begin{align*}
\VField{J_{\pvec{k}_{\mathrm{B}}}}(\pvec{r}) = \sum_{\pvec{l}\in \nnum^2} e^{i\pvec{k}_\mathrm{B}^\mathrm{T} \pvec{a} \cdot \pvec{l}} \VField{J}(\pvec{r}-\pvec{a}\cdot\pvec{l}).
\end{align*}
One readily verifies that $\VField{J_{\pvec{k}_{\mathrm{B}}}}$ is Bloch-periodic with phase vector $\pvec{k}_{\mathrm{B}}$. Hence the corresponding Bloch-periodic solution $\VField{E_{\pvec{k}_{\mathrm{B}}}}$ can be computed on a unit cell. The solution $\VField{E}$ for the orginal source term $\VField{J}$ is then obtained by the inverse Floquet transform. For doing this we introduce the reciprocal lattice vectors $\pvec{b}_1,$ $\pvec{b}_2$ satisfying 
\begin{align*}
\left[ \pvec{b}_1, \pvec{b}_2 \right]^{\mathrm{T}}  \cdot \left[ \pvec{a}_1, \pvec{a}_2 \right] =  2 \pi
\left[
\begin{array}{cc}
1 & 0 \\
0 & 1 
\end{array}
\right],
\end{align*}
with $\pvec{b}_1,$ $\pvec{b}_2$ perpendicular to the plane spanned by $\pvec{a}_1,$ $\pvec{a}_2.$ Again we write $\pvec{b} = [\pvec{b}_1,\,\pvec{b}_2]$ and define the Brillouin zone
\begin{align*}
\mathrm{BZ} = \left\{ \pvec{k}_\mathrm{B}=\pvec{b} \cdot \pvec{\tau} \; | \; \tau \in [0,1]\times [0, 1] \right\}.
\end{align*}  
Integrating $\VField{J_{\pvec{k}_{\mathrm{B}}}}$ over the Brillouin zone reproduces the original source field $\VField{J}:$
\begin{align*}
\int_\mathrm{BZ} \VField{J}_{\pvec{k}_\mathrm{B}} \dd{\pvec{k}_\mathrm{B}} = &  \int_\mathrm{BZ} \sum_{\pvec{l}\in \nnum^2} e^{i\pvec{k}_\mathrm{B}^\mathrm{T} \pvec{a} \cdot \pvec{l}}  \VField{J}(\pvec{r}-\pvec{a} \cdot \pvec{l})\dd{\pvec{k}_\mathrm{B}}, \\ 
= & |\mathrm{BZ}| \sum_{\pvec{l}\in \nnum^2}\VField{J}(\pvec{r}-\pvec{a} \cdot \pvec{l}) \int_{[0,1]\times[0,1]} e^{i\tau^\mathrm{T} \pvec{b}^\mathrm{T} \pvec{a} \cdot \pvec{l}} \dd{\tau}
\end{align*}
Using that $\pvec{b}^\mathrm{T} \pvec{a}=2 \pi \mathbf{I}$ gives 
\begin{align*}
\int_{[0,1]\times[0,1]} e^{i\tau^\mathrm{T} \pvec{b}^\mathrm{T} \pvec{a} \cdot \pvec{l}} \dd{\tau} = 
\left \{ 
\begin{array}{l}
1,\, l_1,l_2=0  \\
0,\,\mbox{otherwise} 
\end{array}
\right .,
\end{align*}
so that 
\begin{align*}
\VField{J} = &\frac{1}{\mathrm{BZ}}\int_\mathrm{BZ} \VField{J}_{\pvec{k}_\mathrm{B}} \dd{\pvec{k}_\mathrm{B}},\;\mbox{and}\,\mbox{consequently} \\
\VField{E} = &\frac{1}{\mathrm{BZ}}\int_\mathrm{BZ} \VField{E}_{\pvec{k}_\mathrm{B}} \dd{\pvec{k}_\mathrm{B}}.
\end{align*}

Since the Fourier transform of $\VField{E}$ is a linear functional, it can be directly gained from the discrete Fourier modes of the contributing Bloch fields $\VField{E}_{\pvec{k}_\mathrm{B}}.$ The same holds also true for the computation of the total emitted power by the dipole source.

\begin{figure}
  \begin{center}
      \includegraphics[width=1.0\textwidth]{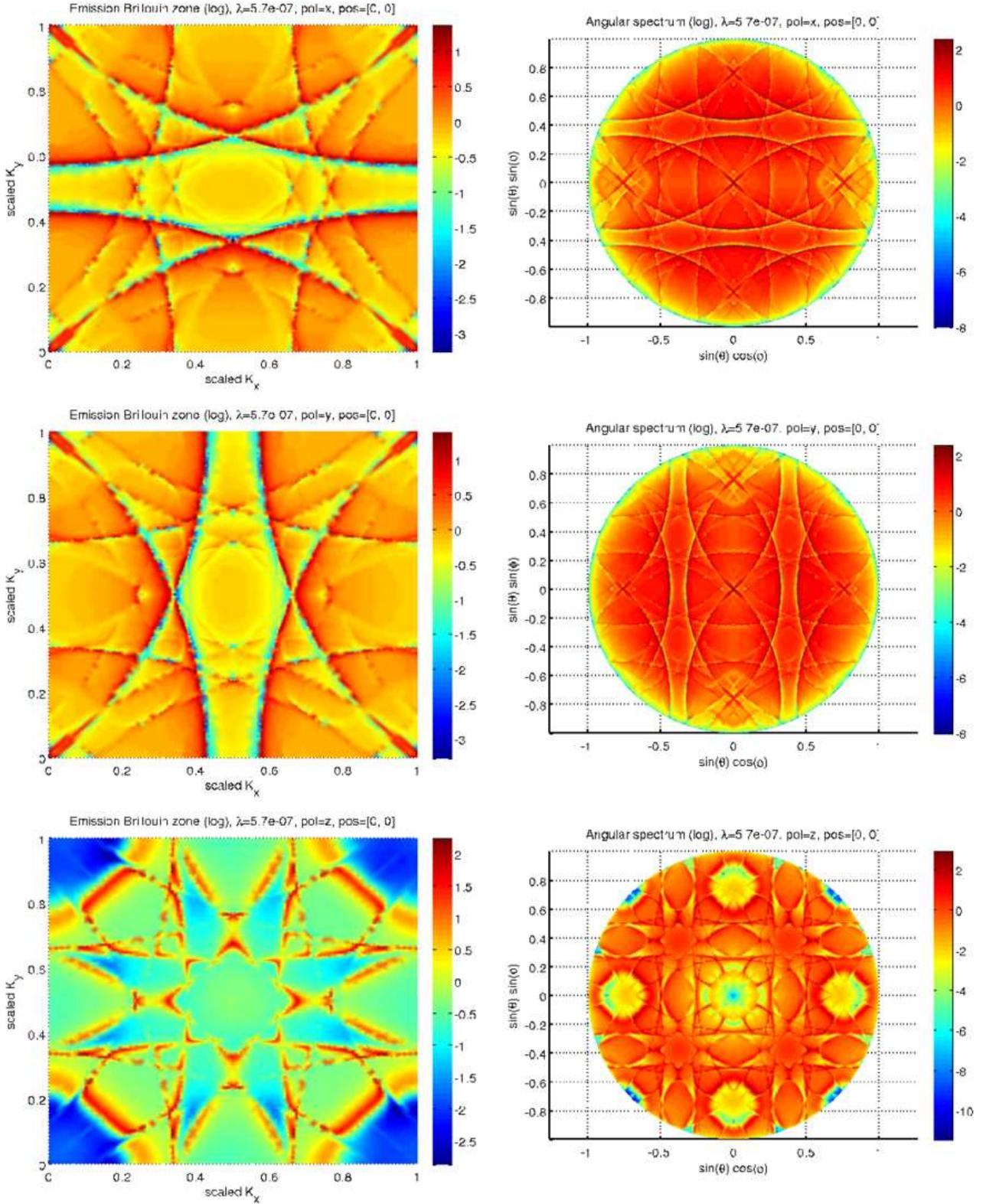}
  \end{center}
  \caption{\label{Fig:EmFar} Emission and far field patterns generated by 
single dipoles ($\lambda=570\mathrm{nm}$, position A, polarizations $x,$ $y,$ $z$).  
Left: Emission of Bloch-periodic dipoles as function of scaled Bloch-phase vector (logarithmic scale). 
The total emission is obtained by integration over the Brillouin zone. 
Right: Far field patterns of the isolated dipoles (radiated intensity as function of emission angle, where 
$\theta=\phi = 0$ corresponds to normal emission, on a logarithmic scale).
}  
\end{figure}

We remark that solving for $\VField{E}_{\pvec{k}_\mathrm{B}}$ is ill-posed, when $\pvec{k}_\mathrm{B}$ corresponds to a Bloch mode. Strictly, these modes do not appear within the Brillouin zone when lossy materials are present. However, the numerical evaluation of the inverse Floquet transform can be heavily affected for  $\pvec{k}_\mathrm{B}$ close to a resonance. This can be cured by introducing a small artificial damping. The so smoothed integral can be efficiently computed by an adaptive integration technique as demonstrated in Pollok et al.~\cite{Pollok:10}. 

\subsubsection*{Remark}
The usage of a small artificial damping resembles the {\em limiting absorption principle} as, for example, discussed for photonic crystals by Joly et al.~\cite{Joly:06a}. There, the case of a fully periodic structure is regarded (threefold periodicity in 3D) and the limiting absorption principle became necessary to single out the correct outward radiating solution. The same is needed here, when the twofold periodic OLED structure supports an undamped Bloch mode which is evanescent in the vertical direction.      

To apply the {\em Brillouin zone integration} technique to an isolated dipole  $\VField{J}(\pvec{r})=\pvec{p}\delta(\pvec{r}-\pvec{r}')$ we have to solve for a Bloch-periodic arrangement of dipoles:
\begin{align*}
\VField{J_{\pvec{k}_{\mathrm{B}}}}(\pvec{r}) = \sum_{\pvec{l}\in \nnum^2} e^{i\pvec{k}_\mathrm{B}^\mathrm{T} \pvec{a} \cdot \pvec{l}} \pvec{p} \VField{\delta}(\pvec{r}-\pvec{r}'-\pvec{a}\cdot\pvec{l}).
\end{align*}
Surely, due to the poor regularity it is also needed to apply the subtraction approach in this case. This can be done by the analytic representation of the Bloch-periodic Green's tensor, see Moroz~\cite{Moroz:06} and the references therein for the Helmholtz equation. As an alternative, we may still use the isolated Green's function as the singular part. Then the correction field is no longer Bloch-periodic but jumps across the periodic boundary of the unit cell,
\begin{align*}
\VField{E}_{c}(\pvec{r}+\pvec{a}  \cdot \pvec{l}) = e^{i\pvec{k}_\mathrm{B}^\mathrm{T} \pvec{a} \cdot \pvec{l}} \VField{E}_{c}(\pvec{r})+ e^{i\pvec{k}_\mathrm{B}^\mathrm{T} \pvec{a} \cdot \pvec{l}} \VField{E}(\pvec{r})_{s}-\VField{E}_{s}(\pvec{r}+\pvec{a}  \cdot \pvec{l}).
\end{align*}  
Fortunately, this jump condition can be seamlessly incorporated in the finite element discretization together with additional Neumann-type boundary terms arising in the variational form.

\begin{table}[t]
\begin{center}
\begin{tabular}{rrrrr}
Position &  $\lambda [\mathrm{nm}]$ & $\eta_{\mathrm{rad, x}}$  & $\eta_{\mathrm{rad}, y}$  & $\eta_{\mathrm{rad}, z}$ \\ \hline 
A & 450 & $0.8756$  & $0.8674$ & $0.5366$\\
B & 450 & $0.8473$  & $0.8407$  & $0.3634$ \\
C & 450 & $0.8330$  & $0.8374$  & $0.2908$ \\
A & 570 & $0.9513$  & $0.9494$ & $0.7001$ \\
B & 570 & $0.9397$  & $0.9024$  & $0.5994$ \\
C & 570 & $0.9149$ & $0.9136$ & $0.4758$ \\
A & 640 & $0.8962$ & $0.8925$ & $0.7395$ \\
B & 640 & $0.9586$ & $0.9427$ & $0.7193$ \\
C & 640 & $0.9386$ & $0.9399$ & $0.7650$ \\
\end{tabular} 
\caption{\label{Tab:Eff} Computed efficiencies for different wavelengths, dipole positions and polarizations. Position A is above the center of a cylinder). Position B is between the first and second cylinder in $x-$direction $x'=500\,\mathrm{nm},\,y'=0.0\,\mathrm{nm}$), and C is the center of 4 neighboring cylinders ($x'=500\,\mathrm{nm},\,y'=500\,\mathrm{nm}$).}                     
\end{center}
\end{table}

\section{Numerical example}
We apply the method to a test case shown in Figure~\ref{Fig:Geo3D}. 
Single dipoles with $x$-, $y$-  and $z$-polarizations are placed at three different positions in the emitter layer 
of a periodically structured 3D OLED setup. 
The periodically arranged scatterers consist of silver cylinders with height $50\,\mathrm{nm}$ 
and diameter $110\,\mathrm{nm}.$ The grid vectors are $a_1=[500; 0; 0]\,\mathrm{nm}$ and  $a_2=[0; 500; 0]\,\mathrm{nm}$.
Figure~\ref{Fig:EmFar} (right) shows the computed far fields for $\lambda=570\,\mathrm{nm}$ and 
position A (above the center of the cylinder, $x'=0.0\,\mathrm{nm},\,y'=0.0\,\mathrm{nm}$). 
On the left hand side the computed Floquet transformed total emitted power $P_{\pvec{k}_\mathrm{B}, \mathrm{tot}}$ is shown. 
The emitted power for the isolated dipole is computed by an integration over the Brillouin zone. 
The sharp structures are due to the presence of complex Bloch-mode resonances near the Brillouin zone. 
To resolve this correctly an adaptive integration technique was applied. 
No artifial damping was used to mollify the integral. 
Table~\ref{Tab:Eff} gives the computed efficiencies for various wavelength and dipole positions. 
The FEM discretization was chosen to guarantee a relative error of $1\%$ in the quantity of 
interest ($\eta_{\mathrm{rad}}$). 
This explains the slight asymmetry in the results 
(differences in $\eta_{\mathrm{rad, x}}$  and $\eta_{\mathrm{rad}, y}$ for positions A and C). 
For generating numerical results we have used the also commercially available FEM solver {\it JCMsuite}. 
                               
In summary, this example demonstrates that FEM based methods can 
accurately simulate electromagnetic near field distributions excited by single emitters in periodically structured 
media. The method can also be applied to arbitrary (non-periodic) structures. 
In post-processes, highly accurate numerical results for
extraction efficiency of radiated power, or other derived quantities of interest can be generated. 
In comparison to supercell methods where very large computational domains have to be used 
(typically beyond 20 $\times$ 20 unit cells) 
the FEM computation on a single unit cell allows for very compact data space requirements and short computation 
times for single computations. 
The numerical integration over the Brillouin-zone can be parallelized 
in a straight-forward manner.

\section*{Acknowledgments} 
This work has been supported by the IM3OLED project
{\it (Integrated multidisciplinary and multiscale modeling for organic light-emitting diodes, NMP-2011.1.4-5/295368)} 
in the Seventh Framework Programme (FP7) of the European Union.
\bibliography{lit}   
\bibliographystyle{spiebib}   

\end{document}

%% file: MatheDef.tex
\newcommand{\nnum}{\mathbb{N}}

\newcommand{\rnum}{{\bf {R}}}
\renewcommand{\Re}{{\mathrm{Re}}}
\renewcommand{\Im}{{\mathrm{Im}}}

\newcommand{\curl}{\nabla \times }

\newcommand{\dd}[1]{\mathrm{d} #1}

\newcommand{\VField}[1]{{\bf #1}} 
 
\newcommand{\pvec}[1]{{\bf{ #1}}}

\def\squareforqed{\hbox{\rlap{$\sqcap$}$\sqcup$}}
\def\qed{\ifmmode\else\unskip\quad\fi\squareforqed}